\documentclass[twocolumn,showpacs,preprintnumbers,amsmath,amssymb]{revtex4}
\usepackage{graphicx}
\usepackage{amssymb}
\usepackage[latin1]{inputenc}
\usepackage{dcolumn}
\usepackage{bm}

\begin{document}



\title{Surface Magnetism in Topological Crystalline Insulators}

\author{Sahinur Reja$^{1}$, H.A.Fertig$^{1}$, L. Brey$^{2}$ and Shixiong Zhang$^1$}
\affiliation{
$^1$Department of Physics, Indiana University, Bloomington, IN 47405\\
$^2$ Instituto de Ciencia de Materiales de Madrid, (CSIC),
Cantoblanco, 28049 Madrid, Spain
}

\date{\today}

\pacs{73.20.At,75.70.Rf,75.30.Gw}

\begin{abstract}
We study topological crystalline insulators doped with magnetic
impurities, in which ferromagnetism at the surface lowers the electronic
energy by spontaneous breaking of a crystalline symmetry.  The number of
energetically
equivalent groundstates is sensitive to the crystalline
symmetry of the surface, as well as the
precise density of electrons
at the surface.  We show that for a SnTe model in the
topological
state, magnetic states can have two-fold, six-fold symmetry, or eight-fold
degenerate minima.
We compute spin stiffnesses within the model to demonstrate the stability
of ferromagnetic states, and consider their ramifications for thermal
disordering.  Possible experimental consequences of
the surface magnetism are discussed.
\end{abstract}
\maketitle
\textit{Introduction} -- Topological crystalline insulators (TCI's) are a class of
materials in which
the energy bands can host non-trivial topology protected by a crystalline symmetry
\cite{Fu_2011}.  These systems support surface states \cite{Liu_2013}
which
remain gapless provided the crystal symmetry is unbroken, and are believed
to present themselves in
(Sn,Pb)Te and related alloys
\cite{Hsieh_2012,Tanaka_2012,Xu_2012,Dziawa_2012,Okada_2013,Yan_2014}.
Interesting effects may arise when the symmetry protecting a topological
band structure is broken.  In
topological insulators protected by time-reversal symmetry (TRS), magnetic impurities
on a surface break this symmetry and form collective states
\cite{Liu_2009,Biswas_2010,Garate_2010,Abanin_2011,Liu_Sau_2016},
which may be understood in terms of a gap opening in the surface spectrum
\cite{Efimkin_2014}.

\begin{figure}[t]
\includegraphics[width=\linewidth]{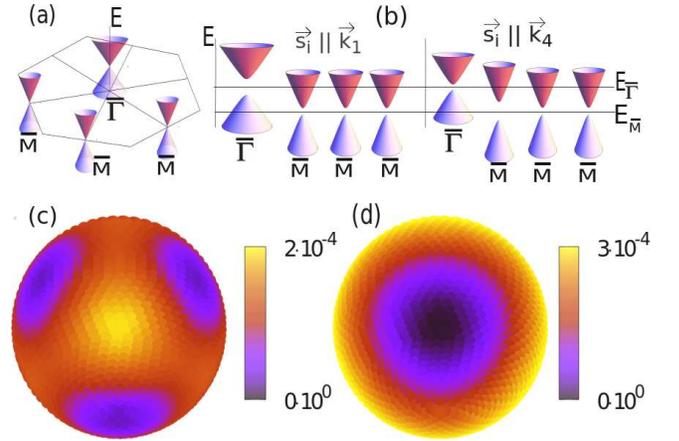}
\caption{(Color online.) (a) Schematic diagram of low energy states on
the (111) surface.  Note the three-fold symmetry.  (b) Gaps induced by
magnetic moments, with relative sizes depending on their orientation.
(c) and (d) Total electronic energy per surface for fixed particle number atom vs. magnetization
orientation
on the Bloch sphere, in units of nearest neighbor hopping $t$.
Only one hemisphere is shown in each figure, with
the ${\bf k}_1$
direction represented by the center. In (c) $\mu \sim E_{\bar M}$; in (d),
$\mu \sim E_{\bar \Gamma}$.}
\label{BlochSphere}
\end{figure}

In contrast, TCI's are not protected by TRS,
so the loss of this symmetry
does not by itself energetically favor ordering of magnetic moments
\cite{Shen_2014,Assaf_2015}.  However,
a uniform magnetization can undermine one or more relevant
crystalline symmetries \cite{FanZhang_2013,Fang_2014}.
Indeed, the most common such symmetry is reflection across
a mirror plane, of which there can be several.
We show below that spontaneous {\it surface} magnetization
opens a maximal gap when oriented
along axes dictated by the {\it bulk} $\,$symmetries of the system.
For a generic surface with a single mirror plane,
there are two surface Dirac points at different momenta and energies \cite{Ando_2015},
and in such cases at low temperature this results in a metallic,
Ising-like ferromagnet,
with the easy axis determined by the chemical potential $\mu$.
Importantly, the number of degenerate low-energy directions is enhanced for surfaces
with further symmetries.  Rotational symmetries in particular
yield multiple mirror planes,
and connect distinct surface Dirac cones to one another, yielding a multiplicity
of easy axis directions.
For sufficiently high symmetry, all the surface
Dirac points may be related by symmetry operations, resulting in a fully
gapped surface spectrum and a large number of
groundstate orientations.

To illustrate this physics, we present detailed
calculations for the (111) surface of (Sn,Pb)Te
\cite{Shixiong_2013,Yan_2014,Taskin_2014,Jie_2014}, using a known
model Hamiltonian \cite{Littlewood_2010,Hsieh_2012}.
The (111) surface states are
characterized
in this system by four surface Dirac points, one at the $\bar{\Gamma}$ point and one
at each of three $\bar{M}$ points \cite{Hsieh_2012}
[Fig. 1(a).]
When the system is doped by
substitutional isolectronic magnetic impurities,
with $\mu$ adjusted to the bulk gap, the absence of free carriers
in the volume suggests that bulk magnetism will not occur.
However, it can be stabilized on the surface when it opens a gap
in the surface spectrum.  Moreover,
the magnitude of
the various gaps are sensitive to the direction of the magnetization [Fig. 1(b)].
Since the gap centers are at different energies, favored magnetization directions
are determined by $\mu$ [Figs. \ref{BlochSphere} (c) and (d)].  When in the vicinity
of the $\bar{\Gamma}$ Dirac point energy, there is a single easy axis, yielding
an Ising ferromagnet.  With $\mu$
near the $\bar{M}$ Dirac point energies, because
these points are connected by a three-fold rotational symmetry,
there is a six-fold degenerate set of
groundstate orientations.  Remarkably, the sensitivity to $\mu$ implies
that the low-energy orientations can be controlled externally by a gate.

The (111) surface is an example of how symmetry leads to a multiplicity of
magnetic groundstates.  This also occurs at
a (001) surface, where
four surface Dirac cones are supported by two distinct mirror planes.
As explained below,
the high symmetry leads to a potential 8-fold degeneracy of groundstate magnetization directions.
Moreover, the energetic coincidence of all the surface Dirac points
allows for the surface states to be fully gapped, yielding insulating behavior
for a range of $\mu$.

The existence of the degenerate magnetic groundstates should be
detectable via the behavior of their domain wall (DW) excitations, which
proliferate at thermal disordering transitions, or can be frozen in
when the system is zero-field cooled.  Furthermore, because DW's connect regions with
different Chern numbers, they necessarily support
bound conducting states \cite{Jackiw_1977}.
Their energetics can also behave rather differently depending upon the placement of
$\mu$ relative to the surface bands.  We discuss possible
effects of the DW's below.

\textit{Bulk Hamiltonian and Surface States} --
Our analysis employs a tight-binding Hamiltonian $H_{bulk}$
for materials in the (Sn,Pb)Te class,
which is a rocksalt structure (fcc lattice). $H_{bulk}$
involves twelve orbitals: for each spin there are $p_x,\,p_y,\,p_z$ states
on each of two
sublattices, labeled $a$ and $b$, with on-site energies $m_{a,b}$
(see Supplementary Material \cite{Supplement}).
The model represents a direct gap semiconductor
with smallest gaps at the $L$ points [${\bf k} =
{\bf k}_1,{\bf k}_2,{\bf k}_3,{\bf k}_4 \equiv
(\frac{\pi}{2}, \frac{\pi}{2}, \frac{\pi}{2})$,
$(-\frac{\pi}{2}, \frac{\pi}{2}, \frac{\pi}{2})$, $(\frac{\pi}{2}, -\frac{\pi}{2},
\frac{\pi}{2})$,
$(\frac{\pi}{2}, \frac{\pi}{2}, -\frac{\pi}{2})$ in units of the inverse nearest
neighbor
separation].
For $\bf k$
precisely at an $L$-point states have well-defined sublattice index (with on-site
energies $m_a$ and $m_b$).
Adjusting $m_b-m_a$ to an appropriate value $m_0$
brings $a$ and $b$ states into energetic coincidence, forming the basis of a Dirac point at
the Fermi energy.

When $m_b-m_a = m_0+m$ with $m<0$, there is a band inversion and
associated nontrivial band topology \cite{Hsieh_2012},
protected in this system by mirror symmetries,
so that surfaces respecting any of them support gapless
states \cite{Liu_2013}.
Low energy forms of these may be constructed \cite{Liu_2010,Silvestrov_2012,Brey_2014},
as we describe for the specific case of the (111) surface
in the Supplementary Material \cite{Supplement}.  For this surface,
there are Dirac points residing at $\bar \Gamma$ and each of the
three $\bar M$ points -- as illustrated in Fig. \ref{BlochSphere} (a) --
with energies $E_{\bar \Gamma}$ and $E_{\bar M}$ respectively.
Note that in this situation the system has a three-fold rotational symmetry,
which maps
the two degenerate states at $E_{\bar \Gamma}$ onto one another, forming
a two-dimensional representation of this rotation group. States at the
three ${\bar M}$ points form a six dimensional
representation. For each Dirac point, approximate explicit forms of the
the wavefunctions may be constructed \cite{Supplement}, which can
be written as eigenstates of a mirror operator $\tilde\sigma_1$
with eigenvalues $\pm 1$.  Projection of $H_{bulk}$ onto these
surface states allows us to construct effective Hamiltonians
in the vicinity of each Dirac point.

\textit{Magnetic Impurities and Surface Hamiltonians} --
It has long been known that metals in the (Sn/Pb)Te
class
\cite{Inoue_1975,Inoue_1977,Inoue_1979,Story_1986,Karczewski_1992,Geist_1996,Geist_1997,Prinz_1999,Lusakowski_2002},
may be doped with magnetic ions which in some circumstances order
ferromagnetically at low temperature.    In these systems
the magnetic ions enter substitutionally for Sn/Pb atoms,
and the coupling of the magnetic moments
with the conduction electrons can be understood rather well
using an $s-d$ model \cite{Dietl_1994},
$H_{sd} = J\sum_i \vec S({\bf r}_i) \cdot \vec s_i$, where $\vec s_i$ represents an
impurity spin at location ${\bf r}_i$ and $\vec S({\bf r}_i)$ is the conduction
electron spin density \cite{Liu_2016}.  We consider the situation where
the chemical potential is in a gap of the bulk spectrum,
so that free carriers are not present
and bulk magnetic ordering is not expected.  In the TCI state, however,
surface electrons couple the magnetic moments of the substitutional
impurities near the surface, and may lead to
ferromagnetism \cite{rosenberg_2012,lasia_2012}.
We model this by assuming magnetic
impurities are present in the system, on one sublattice, near the surface.

The explicit surface wavefunctions
allow us to project the electron spin operators onto surface
states
for the ${\bar{\Gamma}}$ and $\bar{M}$ points.  As discussed in the
Supplementary Material \cite{Supplement},
the spin operators on a single (say, the $a$)
sublattice
for either
the $\bar{\Gamma}$ or an $\bar{M}$ point may be written
\begin{equation}
\vec{S}^{(a)} = {1 \over 4} \left( u_{a}^2 \tilde\sigma_2, u_{a}^2 \tilde\sigma_1,
(u_{a}^2 - v_{a}^2) \tilde\sigma_3 \right).
\label{GammaBarSpins}
\end{equation}
In this expression, $\tilde\sigma_3$ is $2 \times 2$ matrix whose eigenvectors yield
the combinations of $\tilde\sigma_1$ eigenstates with well-defined
eigenvalues under $2\pi/3$ rotations around a bulk $\Gamma-L$ direction
\cite{Supplement}, and $\tilde\sigma_2 = -i\tilde\sigma_3\tilde\sigma_1$.
The quantities $u_a$, $v_a$ are real coefficients involving the tight-binding parameters
 \cite{Supplement}.
We assume the impurity spins ferromagnetically order and treat the Hamiltonian
in mean-field theory; the linear stability of the state against formation
of a spin-density wave can then be checked.
Projecting $H_{sd}$ onto the subspace of surface
states for a Dirac point using Eq. \ref{GammaBarSpins} leads to an
effective Hamiltonian of the form
\begin{equation}
H_{i} \approx E_{i}+\alpha_{i} (q_2-b_2) \tilde\sigma_1 +
\beta_{i} (q_1-b_1) \tilde\sigma_2 + \Delta_i \tilde\sigma_3,
\label{Hsurf_generic}
\end{equation}
where $i$ denotes either $\bar{\Gamma}$ or one of the $\bar{M}$ points, and
$q_{1,2}$ represent wavevector components along the surface.
(Note the relationships between $(q_1,q_2)$ and $(q_x,q_y,q_z)$ depend on
the specific Dirac point $i$.)
As expected on general symmetry grounds, $\alpha_{\bar\Gamma} = \beta_{\bar\Gamma}$,
but $\alpha_{\bar M} \ne \beta_{\bar M}$.
The offsets $b_1$ and $b_2$ are proportional to components of the impurity
magnetization
perpendicular to ${\bf k}_i$,
while $\Delta_i$ is proportional to the component along it.
The resulting spectra,
$
\varepsilon_{i} = E_{i} \pm \sqrt{\alpha_i^2(q_1-b_1)^2 + \beta_i^2(q_2-b_2)^2 +
\Delta_i^2},
$
provides the important observation that when the moments align along a $\Gamma - L$
direction and $\mu \sim E_i$, a gap opens in the corresponding surface spectrum that
lowers its contribution to the total electron energy \cite{Efimkin_2014}.

\textit{Numerical Studies} --
To test this idea we have numerically computed the electronic energy of a TCI slab
with open (111) surfaces, using
the tight-binding model $H_{bulk}$ \cite{Supplement}, and adding an effective
magnetic
field $\vec b$ near the surface on only the $a$ sublattice, in such a way
that their coupling to the two states associated with the surface Dirac cones is
the same. Our tight-binding parameters are adapted from Ref.
\onlinecite{Fulga_2016},
and we have verified the presence
of four surface Dirac points.
(Some technical details are provided in the Supplementary Material \cite{Supplement}.)
{
Initially we consider a slab with primitive unit cell presenting only a single
site of one sublattice on the surface,} and introduce a surface magnetization
as described above.  While this represents a relatively large density of impurities
(relative to experiment),
it captures the correct qualitative physics, and allows us to study a wide
enough slab that the surfaces are effectively decoupled.  As expected from
the above discussion, among the four Dirac cones
the one with largest magnetization projection along its corresponding
$\Gamma$ - $L$ direction develops the largest gap.
Figs. \ref{BlochSphere} (c) and (d) illustrate
the resulting total electronic energy,
for $\mu \sim E_{\bar M}$ in (c) and $\mu \sim E_{\bar \Gamma}$ in (d).
The energy is minimized in the former
case for $\vec s$ along a $\Gamma - L$ direction associated with an $\bar{M}$ point,
while in the latter minimization occurs for $\vec s$ along ${\bf k}_1$.  This leads
to two degenerate minima for
$\mu \sim E_{\bar\Gamma}$ and six
for $\mu \sim E_{\bar M}$.  (Only half of these can be seen in the
figures.)
Analogous results are found when $\mu$ rather than
particle number is fixed.  As expected,
the number of
minimal energy states reflects the symmetry of the surface.

\begin{figure}[t]
\includegraphics[width=\linewidth]{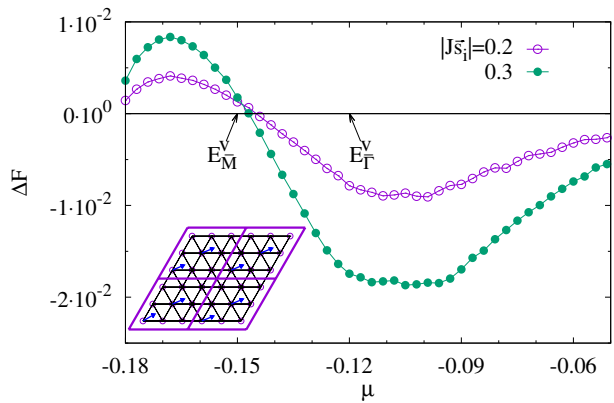}
\caption{(Color online.) Difference in free energy per surface atom
in units of nearest neighbor hopping $t$,
when magnetic moments are oriented in the (111) direction and in the $(11\bar{1})$
direction, as a function of $\mu$,
for two different strengths of $J|\vec s |$ in $H_{sd}$.
$E_{\bar{\Gamma}}^V$ and $E_{\bar{M}}^V$  indicate
valence band tops for $J|\vec s | =0.3$ with $\vec s$ in each direction respectively.
Here there are 2 magnetic ions for 9 atoms on the surface in
the unit cell (inset). }
\label{energy_compare}
\end{figure}

To further substantiate this, we also studied
a more dilute magnetic moment model, in which the impurities
are present only for $2/9$ of the atoms of one sublattice near the surface.
Fig. \ref{energy_compare}
shows the difference in Gibbs free energy of the system
($\langle H_{bulk} \rangle -\mu N$ with $N$ the number of electrons)
when the magnetic moments are oriented in the $(111)$ direction and in the
$(11\bar{1})$
direction, as a function of chemical potential.  The results again demonstrate that
energetically favored directions are determined by $\mu$.
We have also used this geometry to verify
that orienting the two surface magnetic moments in different directions always
raises the energy of the system, supporting our assumed
ferromagnetic
ordering, and that different placements of the impurities in the unit
cell on one of the sublattices has little effect \cite{Supplement}.
The latter suggests
that disorder in the impurity location has little impact on our
results \cite{Liu_2009}.  Finally we note that density of states
resonances \cite{wehling_2014} that appear from vacancies
and impurities in, for example,
graphene, do not appear to be present for the kind of disorder we consider.

The ferromagnetic ordering may be further substantiated by considering
what happens to the electronic energy when the effective field is allowed to
vary spatially with some wavevector $\vec Q$ along an average direction,
$b_{1,2}({\bf r}) = b_{1,2}^{(0)} + \delta b_{1,2} \cos \left({\bf Q} \cdot {\bf
r} \right)$,
$\Delta({\bf r}) \equiv \Delta^{(0)} + \delta \Delta \cos \left({\bf Q} \cdot
{\bf r} \right)$.
To compute this we adopt as our
basic Hamiltonian Eq. \ref{Hsurf_generic}, assuming for simplicity
$\alpha_i =\beta_i\equiv\alpha$, use the directions associated
with $q_1$ and $q_2$ to define $x$ and $y$ directions on the surface,
and compute the change in energy to second order in $\delta b_{1,2}$,
$\delta \Delta$,
and $Q$.
Interestingly, for the valleys
in which there are Fermi surfaces, this turns out
to be independent of $Q$, as is the case for graphene \cite{Brey_2007}.
Thus the spin stiffness (quadratic dependence of the energy correction on $Q$) comes
from any valley(s) for which the Fermi energy passes through a gap.
After an involved calculation (see Supplementary Material \cite{Supplement}),
one finds a correction of the form
\begin{equation}
\frac{\delta E(Q) - \delta E(0)}{\cal S} =
{1 \over 2} \sum_{\mu,\nu=x,y}\rho_{\mu,\nu} Q_{\mu}Q_{\nu},
\label{GradEnergy}
\end{equation}
where the coefficients $\rho_{\mu,\nu}$ are all second order in the deviations
$\delta b_{1,2}$, $\delta \Delta$, and the eigenvalues of the $2 \times 2$ matrix
it represents are positive
\cite{Note_stiff}.
This demonstrates that if the effective field from the surface magnetization
has a spatial oscillation, the resulting energy increases with increasing
oscillation
wavevector, as should be for a ferromagnetically aligned groundstate.  Note that
the stiffnesses $\rho_{\mu,\nu}$ all
diverge as $1/\Delta^{(0)}$ \cite{Supplement}, a property with interesting
consequences to which we will return.

\textit{(001) Surface} --
The (001) surface of SnTe supports gapless states that differ qualitatively from
those of the (111) surface in that they involve valley admixed states \cite{Liu_2013}
in the vicinity of an $\bar{X}$ point of the surface Brillouin zone.
The effective Hamiltonian for such surface states may be written in the form
\begin{equation}
H_{\bar{X}}=v_1q_1\tilde\sigma_1+v_2q_2\tilde\sigma_2+E_{\bar X} +\eta_1\tilde\mu_x\tilde\sigma_1 +\eta_2\tilde\mu_y,
\label{HbarX}
\end{equation}
where $q_1=(q_x+q_y)/\sqrt2$, $q_2=(q_x-q_y)/\sqrt2$, $\tilde\mu_{x,y}$ are a Pauli
matrices acting
on a two-fold valley space, and $\eta_1$, $\eta_2$ are phenomenological
valley-mixing parameters \cite{Liu_2013}.  This expression may be arrived at by
explicit projection of two valleys (e.g., ${\bf k}_1$ and ${\bf k}_4$) onto
(001) surface states using the approach described above.  Dirac points for this
Hamiltonian lie at the momenta ${q}_{1}=\pm (\eta_1^2+\eta_2^2)^{1/2}\equiv q_{\pm}$, $q_2=0$.  Explicit forms for the two zero energy surface states at each of these points
may be obtained using the same approach as for the (111) surface, and effective electron
spin operators on the $j=a$ sublattice derived which couple to magnetic impurities.  Assuming these are ferromagnetically
aligned, in mean-field theory they add a term of the form $\vec{h} \cdot \vec{S}^{(a)}$,
where $\vec{h}$ is the average magnetization, and induce a gap $\Delta$ at the Dirac points satisfying
\begin{eqnarray}
\frac{q_{\pm}^2\Delta^2}{4} = [\tilde D(h_x+h_y)-\sqrt{2}\tilde C(h_x-h_y)q_{\pm}]^2 \nonumber\\
+2[\tilde A\eta_2-\tilde B\eta_1]^2(h_x+h_y)^2
+[\tilde D\eta_1+u_a^2\eta_2/\sqrt{6}]^2h_z^2,\nonumber
\end{eqnarray}
where
$\tilde A=(u_a^2-v_a^2)/12\sqrt{6}+4u_av_a/\sqrt{12}$,
$\tilde B=(u_a^2+2v_a^2-4\sqrt{2}u_av_a$,
$\tilde C=u_a^2/2-v_a^2$, and
$\tilde D=(v_a^2-u_a^2)/\sqrt{12}$.
For fixed magnitude of $\vec h$, it is easy to see that there are two oppositely oriented directions that maximize $\Delta$; moreover, these directions are different for $q_+$
and $q_-$.  Together with the four-fold symmetry that guarantees equivalent behavior
for the coupled $\bf{k}_2-\bf{k}_3$ valleys, we conclude that there are {\it eight}
minimal energy directions for $\vec{h}$ for the (001) surface when the Fermi
energy is in the vicinity of $E_{\bar X}$.
Furthermore, because all the Dirac cones are centered at this energy, we expect
the gaps will generally overlap, so that the chemical potential may pass through
all of them simultaneously.  The spontaneously magnetized (001) surface then allows
for {\it insulating} electronic behavior, in contrast to the (111) surface which
for dilute magnetic impurities
remains metallic.


\textit{Discussion and Speculations} --
We next consider some physical consequences of the surface magnetism
discussed above, focusing on temperature ranges where the impurity magnetic
moments may be treated classically.
As mentioned above, the sensitivity of the magnetization directions for energy minima to $\mu$ should allow it to be controlled via
a gate potential, which in principle would be observable in direct magnetization
measurements.
Another basic observation is that the gap openings induce a Berry's curvature in
the surface bands, which generically induces
an anomalous Hall effect.  For the SnTe system with (111) surfaces
we do not expect it to be quantized
\cite{FanZhang_2013,Fang_2014}, since
the chemical potential typically cannot pass through a gap for all the
surface Dirac species at the same time.  By contrast, for (001) surfaces where
the Dirac cones points are initially all at the same energy, the surface magnetization
may allow all the induced gaps to overlap.  With chemical potential in this
gap, one does expect a quantized anomalous Hall effect
\cite{FanZhang_2013,Fang_2014}.

It is interesting to consider possible consequences of
$\rho_{\mu\nu} \sim 1/\Delta^{(0)}$ as discussed above.
In particular we expect that the multiple minima presented
in Fig. \ref{BlochSphere} imply that there should be
DW excitations in the system, with energy
per unit length scaling as $\sqrt{\Delta^{(0)}\rho_0}$,
with $\rho_0$ an appropriate average of $\rho_{\mu\nu}$'s.
This remains
{\it finite} even as $\Delta^{(0)}$ vanishes, as should happen at high enough
temperature.
The divergence of
the spin stiffness, $\rho_0 \sim 1/\Delta^{(0)}$, reflects the fact that
as the gap vanishes, the quantity $\delta E({\bf Q})$ is no longer analytic
in ${\bf Q}$, and in particular rises {\it linearly} with $Q$ in the long
wavelength limit \cite{Brey_2007}, suggesting a non-local interaction
among spin gradients.  Presuming $\mu$ ends up at the Dirac point as
$\Delta^{(0)}$ vanishes, the simplest model of the system is a clock model
with long-range interactions, which in the Ising case would approach
the transition with mean-field exponents \cite{Paulos_2016}.  For other values of $\mu$
it is possible that DW's with finite energy per unit length can be stabilized
in this circumstance, but if so would not have the simplest
structure \cite{Rajaraman_book}.  In the case of the (111) surface,
for $\mu \sim E_{\bar\Gamma}$
the system presumably will undergo a second order phase
transition in the Ising universality class.  By contrast,
for $\mu \sim E_{\bar M}$ case
with six different minima
the system could be represented by a six-state
clock model. For the (001) surface and $\mu \sim E_{\bar X}$, an
eight-state clock model is relevant.  In both these cases, presuming $\mu$
does not evolve precisely to a Dirac point as the magnetization disorders and
the gaps close,
the phase
transitions should be in the Kosterlitz-Thouless universality
class \cite{Jose_1977}.

We also note that DW excitations in this system
may accumulate charge,
both due to mid-gap states \cite{Jackiw_1977,Schakel_2008}, and from the
surface valleys which
have Fermi surfaces allowing low-energy scattering.
The existence of mid-gap states is a necessary by-product of the topology
of the gapped Dirac points described by Hamiltonians of the form in
Eq. \ref{Hsurf_generic}.  The essential effect of the DW on the electrons
is that $\Delta_i$
changes sign as one moves through it.  This means the Chern number evolves
from $-1/2$ on one side of the DW to $1/2$ on the other,
introducing gapless states bound to the
DW interior \cite{Fang_2014}.
At the critical temperature $T_c$ where the transition occurs, one expects
DW's to proliferate, opening a channel for conduction which is absent below $T_c$.
This could lead to singular behavior (e.g., a cusp) in the conductivity
at the transition \cite{Jungwirth_2001,Dhochak_2015}, and should also have a
signature
when the surface is probed via tunneling.  A further possibility to probe the
physics is by looking for
differences in conductivity between
field-cooling and zero-field cooling of the system
through its critical temperature.
The latter leads to nucleation of groundstate domains with random
orientation, and DW's between them, which cannot relax on the
time scale of an experiment.  Thus one expects stronger surface
conduction from a zero field-cooled sample \cite{Assaf_2015,Ueda_2015,Tian_2016}.
Finally,
the presence of charged DW's on the surface might be detected
directly via coupling to electro-magnetic waves \cite{Ma_2015},
whose scattering should be sensitive to the proliferation of DW's.

In summary, the surface of a magnetically-doped TCI hosts magnetic ordering in
the topological state even when the bulk is disordered.  The unique
electronic structure of a TCI surface leads to a richer set of possible ordered
states
than would be expected from time-reversal symmetry protected topological insulators,
and implies a number of unusual physical behaviors.

\textit{Acknowledgements --} The authors thank Fernando de Juan for helful comments,
and U. Nitzsche for technical assistance.
This work was supported by the NSF through Grant Nos.
DMR-1506263 and DMR-1506460, by the US-Israel Binational Science Foundation, and by
MEyC-Spain under grant FIS2015-64654-P.  Computations were carried out on the
ITF/IFW and IU Karst clusters.  HAF also thanks the Aspen Center for Physics
where some of this work was performed.

\bibliography{MagDopedTCI_references_sahinur}

\end{document}


\title{Supplementary information for ``Surface Magnetism in Topological Crystalline Insulators''}

\author{Sahinur Reja$^{1}$, H.A.Fertig$^{1}$, L. Brey$^{2}$ and Shixiong Zhang$^1$}
\affiliation{
$^1$Department of Physics, Indiana University, Bloomington, IN 47405\\
$^2$ Instituto de Ciencia de Materiales de Madrid, (CSIC),
Cantoblanco, 28049 Madrid, Spain
}

\maketitle


\section{Bulk Hamiltonian}
\label{sec:TBmodel}
Our analysis begins with a tight-binding model \cite{Hsieh_2012}
for materials in the (Sn,Pb)Te class,
which is a rocksalt structure, i.e. an fcc lattice with a two sublattices, such that atomic
species alternate among the points of a simple cubic lattice.  The Hamiltonian of the system
is given by
$H_{bulk} = H_m+H_{nn}+H_{nnn}+H_{so}$, with
\begin{widetext}
\begin{eqnarray}
H_m &=&\sum_j m_j \sum_{{\bf R},s}
\vec{c}^{\,\,\dag}_{j,s}({\bf R}) \cdot \vec{c}_{j,s}({\bf R}) ,
\quad\quad\quad\quad \nonumber \\
H_{nn} &=& t \sum_{({\bf R},{\bf R^{\prime}}),s}
\vec{c}^{\,\,\dag}_{a,s}({\bf R}) \cdot \vec{d}_{{\bf R},{\bf R^{\prime}}}
\vec{d}_{{\bf R},{\bf R^{\prime}}} \cdot \vec{c}_{b,s}({\bf R^{\prime}}) + h.c. , \nonumber\\
H_{nnn} &=& \sum_j t_j^{\prime} \sum_{(({\bf R},{\bf R^{\prime}})),s}
\vec{c}^{\,\,\dag}_{j,s}({\bf R}) \cdot \vec{d}_{{\bf R},{\bf R^{\prime}}}
\vec{d}_{{\bf R},{\bf R^{\prime}}} \cdot \vec{c}_{j,s}({\bf R^{\prime}}) + h.c. , \nonumber\\
H_{so} &=&
i \sum_j \lambda_j \sum_{{\bf R},s,s'} \vec{c}^{\,\,\dag}_{j,s}({\bf R}) \times
\vec{c}_{j,s^{\prime}}({\bf R}) \cdot (\vec \sigma)_{s,s^{\prime}}. \quad\quad
\label{eq_bulk_ham}
\end{eqnarray}
\end{widetext}
In these equations
${\bf R}$ labels the sites of a cubic lattice, $j=a,b$ are the species type
(Sn/Pb or Te),
which have on-site energies $m_{a,b}$,
and $s = \uparrow,\downarrow$ is the electron spin.  The 3-vector of operators
$\vec{c}_{j,s}({\bf R})$ annihilates electrons in $p_x$, $p_y$ and $p_z$ orbitals,
and there is a local spin-orbit coupling strength $\lambda_j$ on each site.
($\vec \sigma$ is the vector of Pauli matrices.)
The vectors $\vec{d}_{{\bf R},{\bf R^{\prime}}}$ are unit vectors pointing from
${\bf R}$ and ${\bf R^{\prime}}$, and, finally, the sum over $({\bf R},{\bf R^{\prime}})$
denotes positions which are nearest neighbors, while
$(({\bf R},{\bf R^{\prime}}))$ denotes next nearest
neighbors.

At an $L$ point, it is possible to diagonalize $H_{bulk}$ analytically.  These eigenvalues
are $E_{0,j}=4t_j^{\prime}-\lambda_j+m_j,$
$E_{\pm,j}=-2t_j^{\prime}+\lambda_j/2 \pm R_j+m_j$,
with $R_j \equiv \sqrt{144t_j^{\prime 2}+24t_j^{\prime}\lambda_j+9\lambda_j^2}/2$,
with $j=a,b$  each of these three values being doubly degenerate.
For large enough $m$, the lowest eigenvalue for the $b$ sites
approaches the highest eigenvalue for the $a$ sites, and we can construct a
projected $4 \times 4$ Hamiltonian representing states with energies
$E \sim E_{+,a} \sim E_{-,b}$ which is valid for ${\bf k}$ near an $L$ point.
This projected Hamiltonian is conveniently written in terms of states
which have well-defined quantum numbers upon $2\pi/3$ rotation around a $\Gamma-L$ direction.
For example, for the $L$ point ${\bf k_1} \equiv (\frac{\pi}{2}, \frac{\pi}{2}, \frac{\pi}{2})$,
we define states
\begin{equation}
|m,s,j\rangle \equiv {1 \over {\sqrt{3}}} \Bigl[ |p_x,j\rangle + w^m |p_y,j\rangle + w^{2m} |p_z,j\rangle \Bigr]
\otimes |s\rangle
\nonumber
\end{equation}
where $|p_{x\,(y,z)},j\rangle$ is a $p_{x\,(y,z)}$-orbital on a site of type $j$,
$|s=\pm\rangle$
is a spin state with quantization axis parallel to ${\bf k_1}$, and $w=e^{2\pi i/3}$.
The relevant eigenstates of the Hamiltonian are then
\begin{eqnarray}
|1,j\rangle &=& -u_jw|m=0,s=+,j\rangle + v_j |m=2,s=-,j\rangle, \nonumber \\
|2,j\rangle &=& u_jw^*|m=0,s=-,j\rangle + v_j |m=1,s=+,j\rangle, \nonumber \\
\label{basis}
\end{eqnarray}
with
\begin{eqnarray}
u_j &=& \frac{\sqrt{2}\lambda_j}{\sqrt{(8t_j^{\prime}+\varepsilon_j)^2+2\lambda_j^2}}, \nonumber\\
v_j &=& \frac{8t_j^{\prime}+\varepsilon_j}{\sqrt{(8t_j^{\prime}+\varepsilon_j)^2+2\lambda_j^2}},
\label{uv}
\end{eqnarray}
$\varepsilon_a = -2t_a^{\prime}+\lambda_a/2+R_a$, and
$\varepsilon_b = -2t_b^{\prime}+\lambda_b/2-R_b$.  Note under a $2\pi/3$ rotation
around the (111) direction, $|1,j\rangle \rightarrow e^{-i\pi/3} |1,j\rangle$
and $|2,j\rangle \rightarrow e^{i\pi/3} |2,j\rangle$.

To proceed, we form the ${\bf k}$-dependent bulk Hamiltonian
$H_b({\bf k})= e^{-i{\bf k} \cdot {\bf R}} H_{bulk} e^{i{\bf k} \cdot {\bf R}}$
which acts on states of definite crystal momentum ${\bf k}$, and project this
into the $4 \times 4$ space defined by the states in Eq. \ref{basis}.  Writing
${\bf k} = {\bf k_1} + {\bf q}$, for small $q$ one finds, after considerable algebra
and up to an overall constant,
\begin{widetext}
\begin{equation}
\bar{H}_1 = Aq_3\tau_x -B[q_1 \tilde\sigma_2 + q_2 \tilde\sigma_1]\tau_y
+[m + C_{12}^{(-)}(q_1^2+q_2^2) +C_3^{(-)} q_3^2]\tau_z +
C^{(+)}_{12}(q_1^2+q_2^2) + C^{(+)}_3 q_3^2.
\label{HbulkL1}
\end{equation}
\end{widetext}
In this expression,
$\tilde\sigma_1 \equiv \frac{\sqrt{3}}{2}\sigma_x - {1 \over 2}\sigma_y$ and
$\tilde\sigma_2 \equiv \frac{\sqrt{3}}{2}\sigma_y + {1 \over 2}\sigma_x$ are $2 \times 2$
matrices acting on the 1,2 indices of the basis states (Eq. \ref{basis}),
and $\tau_x$, $\tau_y$, $\tau_z$ are standard Pauli matrices acting on the sublattice index.
The coefficients in Eq. \ref{HbulkL1} are explicitly given by
$A=-2\sqrt{3}(u_au_b+v_av_b)$, $B=\sqrt{6}(-u_av_b+v_au_b)$,
$C_{12}^{(\pm)}={1 \over 2}[(C_a-D_a/2) \pm (C_b-D_b/2)]$,  and
$C_{3}^{(\pm)}=-{1 \over 2}(D_a \pm D_b)$, with
$C_j=2t^{\prime}_j({1 \over 3} +u_j^2-v_j^2)$ and $D_j = 8t_j^{\prime}/3$.
Finally, the wavevector coordinates $(q_1,\,q_2,\,q_3)$ are given in terms of
the wavevector components of ${\bf q}$ by
\begin{eqnarray}
\left(
\begin{array}{c}
q_1 \nonumber \\
q_2 \nonumber \\
q_3 \nonumber \\
\end{array}
\right)
=
\left(
\begin{array}{c c c}
\frac{1}{\sqrt{2}} & -\frac{1}{\sqrt{2}} & 0 \nonumber \\
\frac{1}{\sqrt{6}} & \frac{1}{\sqrt{6}} & -\frac{2}{\sqrt{3}} \nonumber \\
\frac{1}{\sqrt{3}} & \frac{1}{\sqrt{3}} & \frac{1}{\sqrt{3}} \nonumber
\end{array}
\right)
\left(
\begin{array}{c}
q_x \nonumber \\
q_y \nonumber \\
q_z \nonumber \\
\end{array}
\right).
\end{eqnarray}
Eq. \ref{HbulkL1} is essentially equivalent to previously derived
results based on the symmetry of the underlying Hamiltonian \cite{Hsieh_2012},
but the quadratic terms follow from the microscopic model and play an important
role in determining the coupling to surface magnetic degrees of freedom which are
the focus of this study.
Note that the matrix
$\tilde\sigma_1$ carries out a mirror reflection which is a symmetry of $H_{bulk}$.

Analogous approximate forms for
$H_{bulk}$ near the other three $L$ points can be obtained by
appropriate symmetry operations: the states and effective Hamiltonian in the
vicinity of ${\bf k}_4 \equiv ({\pi \over 2},{\pi \over 2},-{\pi \over 2})$
are related to Eqs. \ref{basis} and \ref{HbulkL1}
by a mirror reflection across the $x-y$ plane, and these quantities for
the other two $L$ points can be constructed by $2\pi/3$ rotations around the (111)
direction from those of {\bf k}$_4$.
For example, for the ${\bf k}_4$ point
we denote the mirror operation by $M_z$, and note the fact that $H_{bulk}(k_x,k_y,-k_z) =
M_z H_{bulk}(k_x,k_y,k_z) M_z^{-1}$.  Writing ${\bf k} = {\bf k_4} + {\bf q}'$, one finds
a long-wavelength form for $H_{bulk}({\bf k})$ in the vicinity of ${\bf k}_4$, $\bar{H}_4$,
which is formally the same form
as Eq. \ref{HbulkL1},
but with the caveat that the $\tilde\sigma_i$ operators
act on the mirror reflected states $|1,j\rangle_4 \equiv
M_z|1,j\rangle$ and $|2,j\rangle_4 \equiv M_z|2,j\rangle$, and
${\bf q} \rightarrow {\bf q}'$ with
\begin{eqnarray}
\left(
\begin{array}{c}
q_1' \nonumber \\
q_2' \nonumber \\
q_3' \nonumber \\
\end{array}
\right)
=
\left(
\begin{array}{c c c}
\frac{1}{\sqrt{2}} & -\frac{1}{\sqrt{2}} & 0 \nonumber \\
\frac{1}{\sqrt{6}} & \frac{1}{\sqrt{6}} & \frac{2}{\sqrt{3}} \nonumber \\
\frac{1}{\sqrt{3}} & \frac{1}{\sqrt{3}} & -\frac{1}{\sqrt{3}} \nonumber
\end{array}
\right)
\left(
\begin{array}{c}
q_x \nonumber \\
q_y \nonumber \\
q_z \nonumber \\
\end{array}
\right).
\end{eqnarray}
Rewriting $\bar{H}_4$ in terms of ${\bf q}$, one finds
\begin{widetext}
\begin{eqnarray}
\bar{H}_4=A\left(\eta q_2+q_3\right)\tau_x&-&B\left[ q_1\tilde\sigma_2^{\prime}
+\left(q_2+\eta q_3 \right) \tilde\sigma_1^{\prime} \right] \tau_y
+\left\{ m + C_{12}^{(-)} \left[ q_1^2 +\left(q_2+\eta q_3 \right)^2 \right]
+C_3^{(-)} \left(q_2+\eta q_3 \right)^2 \right\} \tau_z \nonumber \\
&+& C_{12}^{(+)} \left[ q_1^2 +\left(q_2+\eta q_3 \right)^2 \right]
+C_3^{(+)} \left(q_2+\eta q_3 \right)^2,
\label{HbulkL4}
\end{eqnarray}
\end{widetext}
with $\eta=2\sqrt{2}/3$.
$\bar{H}_1$ and $\bar{H}_4$ may be used to construct surface states and effective surface Hamiltonians, as described
in the next section.

\section{Surface Hamiltonians}

As discussed in the main text, when the mass of the system is negative $(m < 0)$ the energy bands
described by $\bar{H}_1$ near the ${\bf k}_1$ point are topological, and we expect that
surfaces will host gapless states \cite{Liu_2013}.
These must be eigenstates of $\bar{H}_1$, $\bar{H}_4$, or of one of the Hamiltonians associated with
${\bf k}_2$ or ${\bf k}_3$, but with $q_3 \rightarrow i\kappa$ in order to make the
state evanescent in the bulk.
Focusing on ${\bf k}_1$, we
set $q_1=q_2=0$
to find surface states exactly at the $\bar{\Gamma}$
point. The evanescent state must satisfy
\begin{equation}
\left[i\kappa A \tau_x +\left(m-C_3^{(-)}\kappa^2 \right) \tau_z \right]
|\kappa\rangle
= \left( E + C_3^{(+)} \kappa^2 \right) |\kappa\rangle.
\label{DPstates}
\end{equation}
Note that arriving at this equation is only possible because we
have retained quadratic terms in $\bar{H}_1$ \cite{Liu_2010,Silvestrov_2012,Brey_2014}.
In order to satisfy Eq. \ref{DPstates},
$\kappa$ must obey
\begin{equation}
\left( E+C_3^{(+)} \right)^2 = \left(m-C_3^{(-)}\kappa^2 \right)^2
-A^2\kappa^2.
\label{E_kappa_condition}
\end{equation}
For a given value of $E$, one finds two values of $\kappa^2$ that satisfy
Eq. \ref{E_kappa_condition}, $\kappa_{\pm}^2$.  With the proper choice of overall
sign for $\kappa_{\pm}$,
the eigenstates $|\kappa_+\rangle$ and $|\kappa_-\rangle$ vanish in real space deep in the
bulk of the system, but generically neither vanishes at the surface.  However if
$|\kappa_+\rangle \propto |\kappa_-\rangle$, we can form a linear combination
of these that does \cite{Liu_2010,Silvestrov_2012,Brey_2014}.
One may show that for
sgn$(mC_3^{(-)})<0$,
there is a unique value of $E$
for which this is possible, given by $E=E_{\bar{\Gamma}} \equiv
-\frac{C_3^{(+)}}{C_3^{(-)}}m$.  Because the states
$| \kappa_{\pm}\rangle$ have no dependence on the $\tilde\sigma_i$ operators
(see Eq. \ref{DPstates}), we
can form two independent pairs of these.


The Dirac point states for the $\bar{M}$ points is obtained in a very analogous way.
For example, starting with $\bar{H}_4$ (Eq. \ref{HbulkL4}),
we make the replacement $q_3 \rightarrow i\kappa$ and set
$q_1=q_2=0$.  One may find two different values of $\kappa$ with the same eigenvectors
of $\bar{H}_4$ at the energy $E_{\bar M} = -m\frac{C_{12}^{(+)}\eta^2+C_3^{(+)}}
{C_{12}^{(-)}\eta^2+C_3^{(-)}}$.  Thus we can construct two states meeting the boundary
conditions.
From the rotational symmetry,
each of the other ${\bar M}$ points will host a pair of
degenerate states at precisely this energy, yielding three sets of Dirac
points.  Note that $E_{\bar M} \ne E_{\bar \Gamma}$, so the Dirac points are at
two different energies, in agreement with numerical tight-binding studies \cite{Polley_2014}.

To simplify subsequent discussion, we drop the particle-hole symmetry-breaking
terms (i.e., set $C_{12}^{(+)},\,C_{3}^{(+)} =0$) in $\bar{H}_1$, $\bar{H}_4$, and the long-wavelength Hamiltonians for the other
$L$ points, except to note that the surface Dirac point energies at the $\bar{\Gamma}$ and
$\bar{M}$ points are different, so that we add $E_{\bar{\Gamma}}$ and $E_{\bar{M}}$ to
to $\bar{H}_1$ and $\bar{H}_4$, respectively.
Thus we make the replacements
\begin{widetext}
\begin{eqnarray}
\bar{H}_1 &\rightarrow& A q_3\tau_x -B[q_1 \tilde\sigma_2 + q_2 \tilde\sigma_1]\tau_y
+[m + C_{12}^{(-)}(q_1^2+q_2^2) +C_3^{(-)} q_3^2]\tau_z +
E_{\bar{\Gamma}}, \nonumber \\
\bar{H}_4 &\rightarrow& A\left(\eta q_2+q_3\right)\tau_x-B\left[ q_1\tilde\sigma_2^{\prime}
+\left(q_2+\eta q_3 \right) \tilde\sigma_1^{\prime} \right] \tau_y
+\left\{ m + C_{12}^{(-)} \left[ q_1^2 +\left(q_2+\eta q_3 \right)^2 \right]
+C_3^{(-)} \left(q_2+\eta q_3 \right)^2 \right\} \tau_z + E_{\bar{M}}. \nonumber \\
\label{H1H4Symmetrized}
\end{eqnarray}
These Hamiltonians are next projected into space of surface states derived above.
The explicit form of these that are relevant for the $\bar{\Gamma}$ point (with effective
Hamiltonian $\bar{H}_1$) are
\begin{eqnarray}
|u_1\rangle &=& {1 \over {\sqrt{2}}}
\left(
\begin{array}{c}
1 \nonumber \\
0 \nonumber \\
-i sgn(A) \nonumber \\
0 \nonumber
\end{array}
\right)
{\cal N}_z\left(e^{-\kappa_+ z} - e^{-\kappa_- z} \right), \nonumber \\
|u_2\rangle &=& {1 \over {\sqrt{2}}}
\left(
\begin{array}{c}
0 \nonumber \\
1 \nonumber \\
0 \nonumber \\
-i sgn(A) \nonumber
\end{array}
\right)
{\cal N}_z\left(e^{-\kappa_+ z} - e^{-\kappa_- z} \right).
\end{eqnarray}

\end{widetext}

\begin{figure}[t]
\includegraphics[width=0.9\linewidth]{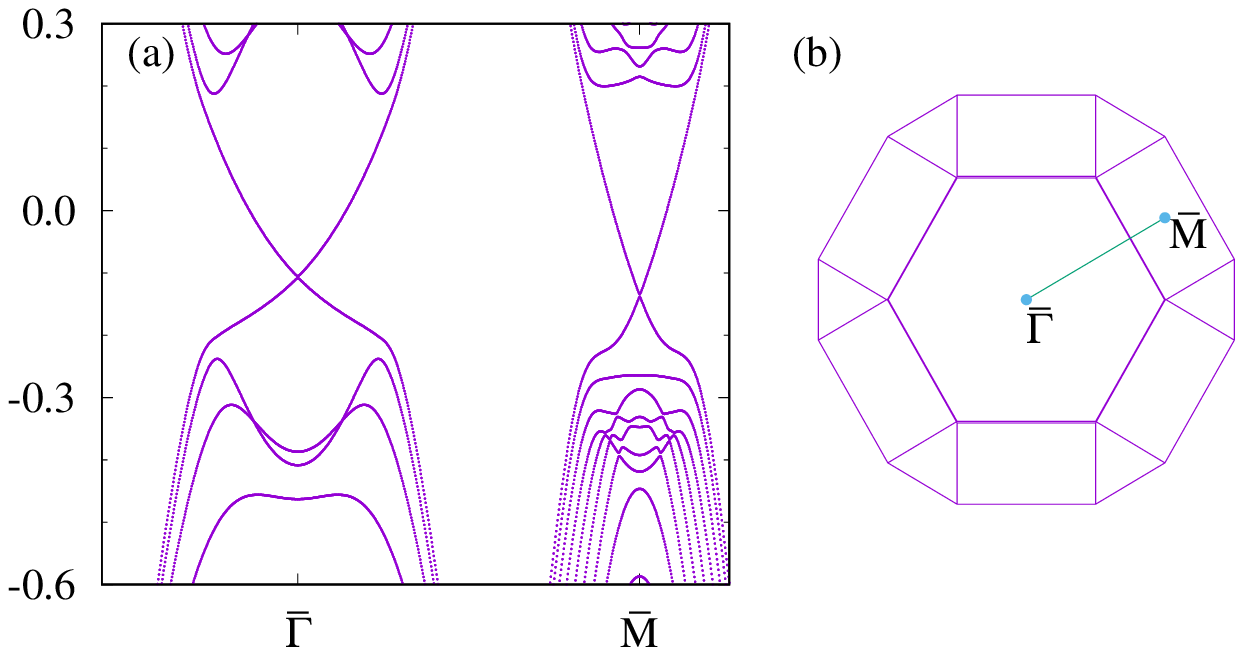}
\caption{(Color online.) The Bnad structure along $\bar{\Gamma}$ to $\bar{M}$ points showing the Dirac cones on (111) suface. The parameter valuse are mentioned in the text.}
\label{fig_dirac_cone_suppli}
\end{figure}

\begin{figure}[t]
\includegraphics[width=0.9\linewidth]{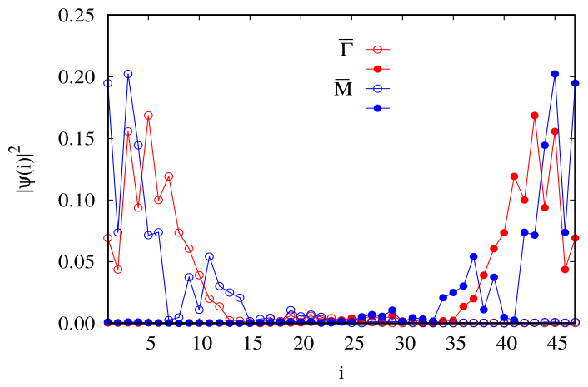}
\caption{(Color online.) Surface state wave function weights $|\psi(i)|^2$ as a function of layer indexed by $i$, stacked along (111) direction. The results are shown for a slab with 47 layers. The red (blue) 
circles indicate states associated with the $\bar{\Gamma}$($\bar{M}$) point of the surface Brillouin zone.
Open (closed) circles indicate states associated with the top (bottom) surface of the slab.
}
\label{surface_state_diff_suppli}
\end{figure}

In these expressions, the four entries are coefficients of the
$(|1,a\rangle, |2,a\rangle, |1,b\rangle, |2,b\rangle)$ states at the ${\bf k}_1$ point,
$z$ denotes the direction perpendicular to the surface (with $z>0$ being points inside the system),
${\cal N}_z$ is a normalization constant, and $\kappa_{\pm}^2=
m + \left[ A^2 \pm \sqrt{4mC^{(-)}_3A^2 + A^4} \right]/2C^{(-)}_3$.
Note that the wavefunctions have equal weight on the two sublattices, and moreover
are eigenstates of the $\tilde\sigma_3$ operator.  The corresponding results for
the ${\bf k}_4$ point may also be written as eigenstates of the relevant $\tilde\sigma_3$
operator (i.e., acting on states near the ${\bf k}_4$ point), and
also have equal weight on the two sublattices, although the relative phase is more
complicated and has the form
$$
e^{i\theta_4} \equiv \frac{A-i\eta B}{\sqrt{A^2 +\eta^2B^2}}.
$$
With the explicit surface wavefunctions in hand
it is straightforward to project $\bar{H}_1$ and $\bar{H}_4$ onto the space of surface states, yielding
surface Hamiltonians for the $\bar{\Gamma}$ and $\bar{M}$ points.  These take the form
\begin{equation}
H_{\bar \Gamma} \equiv B\left[q_1 \tilde\sigma_2 + q_2 \tilde\sigma_1\right]+E_{\bar\Gamma}
\label{HGammaBar}
\end{equation}
and
\begin{equation}
H_{\bar{M}} \equiv \frac{AB}{\sqrt{A^2+\eta^2B^2}} \left[
\left(\eta^2-1\right)q_2 \tilde\sigma_1
+ q_1 \tilde\sigma_2 \right]+E_{\bar{M}}.
\label{HMBar4}
\end{equation}
Note that the states which the $\tilde\sigma_i$ operators act
upon in $H_{\bar{M}}$ are different than those of $H_{\bar \Gamma}$.
Analogous results may be obtained for the other two $\bar{M}$ points by $2\pi/3$
rotations of Eq. \ref{HMBar4}.

As described in the main text,
we assume that there are impurity spins which order
ferromagnetically on the surface, and treat these
in mean-field theory, so that they form an effective uniform Zeeman field
that couples to the electron spin operators, projected onto the surface states,
as shown in Eq. 1 in the main text.
Adding these to the various Hamiltonians (e.g., Eqs. \ref{HGammaBar} and \ref{HMBar4}) leads
to the generic Hamiltonian in Eq. 2 of the main text.

\section{Numerical Methods}
We numerically simulate the model Hamiltonian given in Eq. \ref{eq_bulk_ham} for
(Sn,Pb)Te materials.
As we mentioned above, the lattice structure is fcc with two sublattices $a$ and $b$
(i.e., a rocksalt structure). Our numerical studies focus on the (111) surface,
so that it is convenient to view the fcc lattice as stacked, two-dimensional triangular
lattices, i.e., a close packed structure, with ABC stacking.
Triangular layers of the $a$ and $b$ sublattices are arranged alternately along the stacking direction.

The electronic structure calculations we report are done by taking a finite but relatively large number of
these layers along (111) direction  (i.e., a slab geometry). We used $47$ layers which we found
to be
sufficient to avoid interaction of states of the two slab surfaces. 
The model parameters we used 
in Eq. \ref{eq_bulk_ham}
for the surface state calculations are $t=0.9, t_a^{'}=-t_b^{'}=-0.5,
\lambda_a=\lambda_b=-0.7, m_a=-m_b=3.5$.
These place the system in the topological regime,
as can be seen from the band structure of the slab, shown in Fig. \ref{fig_dirac_cone_suppli},
which demonstrate the presence of gapless states in the bulk gap
around the
$\bar{\Gamma}$ and $\bar{M}$ points in the surface Brillouin zone. 
It is important to note that the energy
of the Dirac point at $\bar{\Gamma}$ is slightly higher than that of $\bar{M}$. 
This asymmetry is responsible for some of the unusual magnetic behaviors described in the main text.

The confinement of wavefunctions with energies in the bulk gap to the surfaces can
be demonstrated explicitly.  If we consider states with crystal momenta precisely
at the $\bar{\Gamma}$ and $\bar{M})$ points, one finds two pairs of states
for each, with each pair associated with one of the two slab surfaces.
Fig. \ref{surface_state_diff_suppli} shows the electron density associated with representative
states for each of the high symmetry points on the surfaces.
From direct examination we see that the surface states penetrate approximately 10
layers into the system bulk for the parameters we use.

As mentioned in the main text, we have modeled the coupling of the electrons to magnetic impurities by an $s-d$ model. 
In real systems these will be randomly located (on a single sublattice)
at different depths from the surface, but on average the coupling of the spins
to both states should be the same for both the $\bar{\Gamma}$ and $\bar{M}$ points.
To model this,
we use the information of the wave functions shown in 
Fig. \ref{surface_state_diff_suppli} by putting magnetic 
moments on {\it two} layers, say $L_1, L_2$, near each surface.
We then choose effective local Zeeman fields on each layer of magnitudes
$(J|\vec s|)_{L_1}$ and $(J|\vec s|)_{L_2}$ such that they satisfy the equation
\begin{equation}
\frac{(J|\vec s|)_{L_1}}{(J|\vec s|)_{L_2}}=
\frac{|\psi_{\bar{\Gamma}}(L_2)|^2-|\psi_{\bar{M}}(L_2)|^2}
{|\psi_{\bar{M}}(L_1)|^2-|\psi_{\bar{\Gamma}}(L_1)|^2}.
\end{equation}
This guarantees that the net coupling of the impurity spins to electron spins
in the $\bar{\Gamma}$
and $\bar{M}$ states will have the same overall strength.
In practice, in our calculations the magnetic impurities reside
 on layer 3 and 5 near one surface and on layer 43 and 45 near the other.

\begin{figure}[t]
\includegraphics[width=0.9\linewidth]{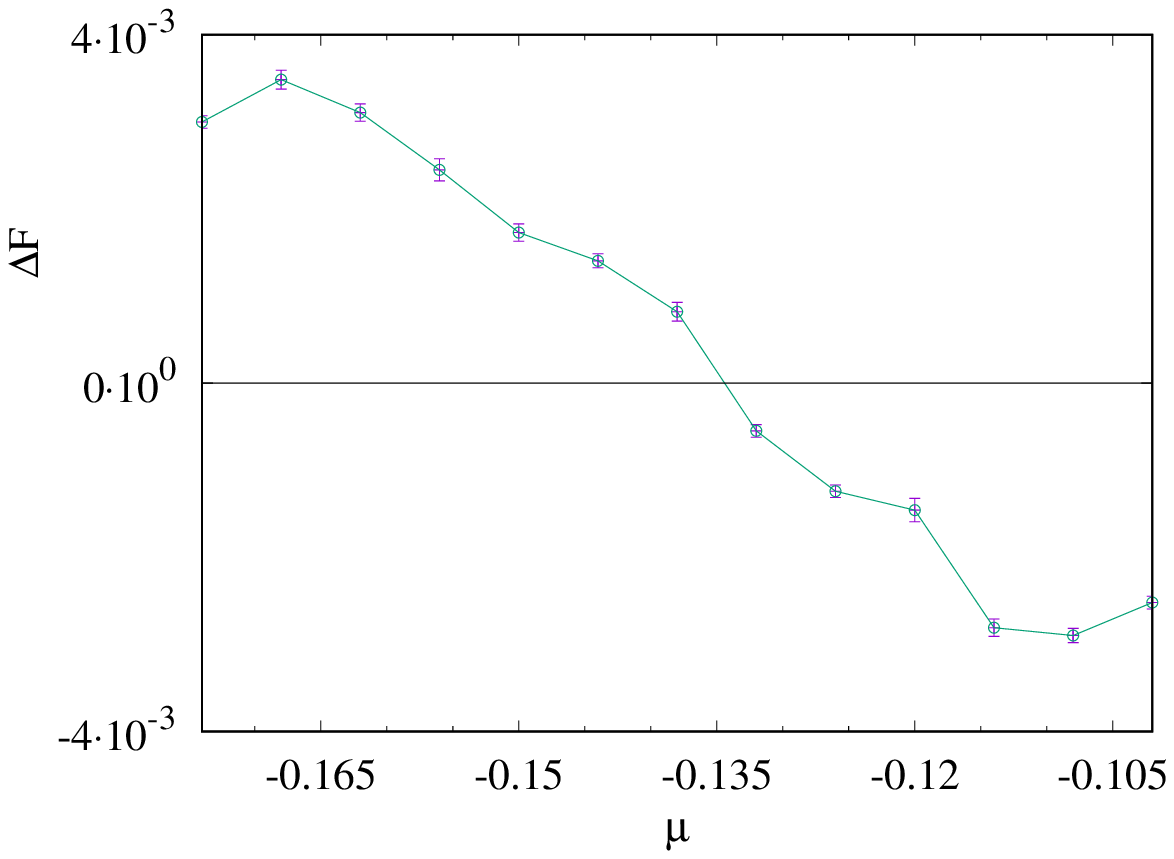}
\caption{(Color online.) Average difference in free energy per surface atom
[in units of $t$] as a function of chemical potential for $J|\vec s |=0.2$
when magnetic moments are oriented in (111) and $(11\bar{1})$
directions.  Results are for 30
random locations of two impurities on {\it a-}sublattice sites,
within five layers of the surface.
The average difference is qualitatively similar to the array of impurities
discussed in the main text. Standard deviation
is shown as error bars.}
\label{energy_diff_suppli}
\end{figure}

\section{Random impurity coupling to surface electrons}
In the main text we discussed the physics when the coupling of the magnetic impurities to surface electron is uniform. But in real materials, this coupling could be randomly distributed, particularly because the magnetic impurities are on random locations among the {\it a-}sublattice sites. To check the effect of this randomness, we simulate the system 30 times, putting two magnetic impurities (dilute limit) at two random {\it a-}sublattice sites within five layers of the system surface. (The top layer of the supercell is illustrated in the inset of Fig. 2
of the main text.) Then we calculate the average and standard deviation for the difference in free energy as a function of chemical potential when magnetic moments are aligned parallel to
the $111$ and $11\bar{1}$ directions respectively.  Fig.\ref{energy_diff_suppli} shows that the energy difference goes from positive to negative in a manner and at a scale very similar to the
behavior illustrated in the main panel of the Fig. 2 of the main text.
This suggests that disorder in the coupling between the magnetic impurities and
the surface electrons does not change the qualitative behavior of the physics of this system.

\section{Spin Stiffness Coefficients}

In this section, we describe our computation of the spin stiffness for a fully filled
band.  For simplicity we consider an isotropic surface Dirac cone, as is found for the
$\bar{\Gamma}$ point; anisotropic Dirac cones ($\bar{M}$ points) should yield
qualitatively similar results.  Our generic Hamiltonian has the form
$$
H_0=\alpha\left[(q_x+b_x)\sigma_x + (q_y+b_y)\sigma_y + b_z\sigma_z\right],
$$
where for simplicity of notation we have rewritten the parameter $\Delta_i$ in the text as $\alpha b_z$.
Eigenstates of $H_0$ are spinors of the form
\begin{equation}
\left(
\begin{array}{c}
u \nonumber \\
v \nonumber
\end{array}
\right)_{\!\!p}
=
\frac{e^{i{\bf q} \cdot {\bf r}}}{{\cal S}^{1/2}
\sqrt{q'^2 +(p\varepsilon_0({\bf q})-b_z)^2}}
\left(
\begin{array}{c}
q_x'-iq_y' \nonumber \\
p\varepsilon_0({\bf q}) -b_z \nonumber \\
\end{array}
\right)
\equiv |{\bf q},p\rangle,
\end{equation}
where $p = \pm 1$ is the band index for states with energy $p\varepsilon({\bf q}) =
p\sqrt{q_x'^2+q_y'^2+b_z^2}$, $(q_x',q_y')=(q_x+b_x,q_y+b_y)$, and ${\cal S}$ is the
area of the surface.  We consider the effect on the total electronic energy of
a filled lower band ($p=-1$) of adding a spatially varying component to ${\bf b}$,
of the form $h=\delta {\vec b} \cdot \vec\sigma \cos {\bf Q}\cdot {\bf r}$ to $H_0$,
focusing on its dependence on ${\bf Q}$ for small $Q$.  Using second order perturbation
theory, one finds for the energy shift $\delta E({\bf Q})$ that

\begin{widetext}
\begin{eqnarray}
\delta E({\bf Q}) - \delta E(0) &\approx&
{1 \over {32}} \sum_{\mu,\nu =x,y} Q_{\mu}Q_{\nu}\sum_{{\bf q}} \left\{
\frac{|\langle {\bf q},-1| \delta {\vec b} \cdot \vec\sigma |{\bf q},+1\rangle|^2}{\varepsilon_0({\bf q})^2}
\partial_{\mu}\partial_{\nu} \varepsilon_0({\bf q})
-\frac{1}{\varepsilon_0({\bf q})} \partial_{\mu}\partial_{\nu}|\langle {\bf q},-1| \delta {\vec b} \cdot \vec\sigma |{\bf q},+1\rangle|^2
\right\} \nonumber \\
&\equiv& \frac{{\cal S}}{2}\sum_{\mu,\nu =x,y} \rho_{\mu,\nu} Q_{\mu}Q_{\nu}.
\end{eqnarray}
With a lengthy albeit in principle straightforward calculation, the coefficients $\rho_{\mu,\nu}$
may all be computed explicitly, with the results

\begin{eqnarray}
\rho_{xx} &=&
\frac{2}{\pi \Delta^{(0)}}
\left[
\frac{2}{5}\, \alpha^2\delta b_x^2 + \frac{8}{15}\, \alpha^2\delta b_y^2 + \frac{4}{15}\, \alpha^2\delta b_z^2
\right], \nonumber\\
\rho_{yy} &=&
\frac{2}{\pi \Delta^{(0)}}
\left[
\frac{8}{15}\, \alpha^2\delta b_x^2 + \frac{2}{5}\, \alpha^2\delta b_y^2 + \frac{4}{15}\, \alpha^2\delta b_z^2
\right], \nonumber\\
\rho_{xy} &=&
\frac{2}{\pi \Delta^{(0)}}
\left[
\frac{8}{15}\, \alpha^2\delta b_x\,\delta b_y
\right], \nonumber
\end{eqnarray}
where $\Delta^{(0)}=\alpha b_z$ is the gap for the unperturbed spectrum of $H_0$.
\hfil\break
\end{widetext}

\bibliography{MagDopedTCI_references}